\documentstyle[aps,eqsecnum,preprint]{revtex}
\begin{document}
\preprint{
\rightline{\vbox{\hbox{\rightline{MSUCL-1115}}}} 
        }
\title{Revisiting lepton pairs at the SPS}

\author{J. Murray}
\address{
Department of Physics, Linfield College, McMinnville, OR 97128}

\author{W. Bauer}
\address{
National Superconducting Cyclotron Laboratory, Michigan State University, 
East Lansing, MI 48824}

\author{K. Haglin}
\address{
Department of Physics and Astronomy, St. Cloud State 
University, St. Cloud, MN 56301}
\maketitle

\begin{abstract}

We confirm the importance of standard medium effects (hadronic rescattering) in
heavy-ion collisions by using a pQCD-based model to investigate the 
dilepton spectra from 
Pb+Au collisions at 158 $GeV$/$nucleon$.  These same effects, namely prompt 
$\pi \rho \to \pi e^+e^-$, have been studied in several CERN SPS
systems\cite{me1}. In addition, the contribution from $\eta$'s
produced by intermediate-stage scattering of pions and previously 
unscattered projectile 
nucleons to the dilepton spectrum has been included. The
results presented here are consistent with previous studies stating that this
type of rescattering effect explains a portion of the ``excess'' lepton pairs
seen by the CERES experiment, but not the entire effect.

\end{abstract}
\pacs{PACS numbers: 25.75.-q, 12.38.Mh, 24.85.+p, 24.10.Lx}

\narrowtext

In future heavy-ion programs, considerable effort will be spent on gaining
information about the space-time history of the collision. This information
becomes increasingly important if a phase transition to quark-gluon degrees of
freedom occurs during the reaction. In order to have an understanding of this
transition, as well as being able to determine its existence, it is necessary
to find signals that will remain intact during all stages of the collision. 
Although hadrons are abundant in the final state, they are only truly
sensitive to the system dynamics after hadronization occurs. In addition, the
resulting hadronic environment will participate in multiple interactions before
reaching the detector. Therefore, reconstructing any information that hadrons
might contain about the initial stages of the collision would be a daunting
task. 

This is the reason electromagnetic signals, such as dileptons and direct
photons, have gained in popularity. These probes, due to a large mean free path
in hadronic matter, appear in the detector after almost no interaction with the
medium. This
property makes electromagnetic signals ideal for studying the early stages of a
heavy-ion collision where this type of phase transition might occur. 
This is why understanding these probes in current heavy-ion experiments is 
essential. In order to shed some light on the higher temperature and density
experiments where ``new'' physics should occur, one must fully understand how
electromagnetic signals behave in lower temperature and density systems.  

With this in mind, we revisit dilepton data taken by the CERES
collaboration\cite{ceres}. As in the S+Au lepton pair measurements, the
invariant mass spectra of dileptons from a Pb projectile, with an energy of 158
$GeV$/$nucleon$, incident on a Au target reported an excess of dileptons over
the collaboration's ``cocktail'' predictions\cite{ceres}. 
Since purely conventional explanations for the excess seem to be
insufficient\cite{drees}, the nature of the enhancement suggests 
several possibilities. Among the most prominent are studies into medium
modifications resulting in a shifted rho mass\cite{rhoshift,both} and 
consequences arising from modifications in the $\pi^+\pi^-\to e^+e^-$ 
reaction\cite{rapp,both}. The study in this paper will be restricted to a more
conventional approach\cite{first_suggestion,haglin,me1} using non-resonant 
scattering of pions and rhos to partially explain the enhancement of 
electron-positron pairs.

To describe the initial stages of an ultra-relativistic heavy-ion collision, it
is necessary to use parton degrees of freedom. With this in mind, there are
efforts underway to construct so-called parton cascades\cite{geiger1,geiger2}.
These
models are based on perturbative QCD and are therefore attractive candidates
for a space-time transport theory in this energy regime. However, we have shown
that there are severe problems with causality violations\cite{caus} and
with the time-ordering of soft-gluon emission\cite{kort}.
        
Therefore, under these circumstances, a much simpler approach might provide
more reliable results: geometrical folding of the results of event generators
for the elementary processes. This prescription is followed, for example, in
HIJING\cite{hijing}. The simulation used in this study is similar to HIJING. 
It employs pQCD and parton distribution functions to characterize the 
individual nucleon-nucleon collisions and uses Glauber-type
geometry\cite{ncoll} to determine the scaling. 
The kinematics of the nucleon-nucleon collisions are handled by PYTHIA
and JETSET\cite{sjostrand}, high energy event-generators using 
pQCD matrix elements as well as the Lund fragmentation scheme. We refer the
reader to our previous work for a more detailed description of the
model\cite{me1}.

Dileptons from pseudoscalars ($\pi^0$, $\eta$, $\eta^{\prime}$) and vectors
($\omega$, $\rho^0$, $\phi$) produced in the primary
scattering phase are not enough to account for the hadron-induced data measured
by the CERES collaboration. 
Therefore, in addition to this type of lepton pairs, our model also 
incorporates secondary scattering of hadronic resonances. All
pions and rhos formed during the primary collisions of nucleons will 
have a chance to scatter amongst themselves before
decaying. The reactions we consider are of two types, one which produces a
resonance that decays to dileptons and the other which goes to dileptons
directly.  

Of the first type, $\pi^+ \pi^- \to \rho^0$ $\to e^+
e^-$ and $\pi^0 \rho^{\pm} \to {a_1}^{\pm}$ $\to \pi^{\pm} 
e^+ e^-$ have been included. Of the second type, 
$\pi^0 \rho^{\pm} \to \pi^{\pm} e^+e^-$ 
has been included and the other isospin channels 
($\pi^{\pm} \rho^{0} \to \pi^{\pm} e^+e^-$, $\pi^{\mp} \rho^{\pm} \to 
\pi^{0} e^+e^-$) are assumed to be of the same magnitude.
To accomplish these types of scattering, pions and 
rhos must of course appear in the final state of the model described in the
previous section. As the default, JETSET automatically decays all hadronic 
resonances, but it also contains provisions to prohibit them.
We thus allow pions and rhos to scatter when
conditions are favorable.  Technically, the steps involved in secondary 
scattering are similar to those for primary (nucleon-nucleon) scattering. 

Since the S+Au dilepton study was published\cite{me1}, several additions have 
been made to the simulation.
Previously, the only pions and rhos that were allowed to secondary scatter came
from the string fragmentation stage of the simulation. Pions and rhos resulting
from a hadronic decay chain were prevented from rescattering. These particles
are now allowed to scatter and contribute to the dilepton spectra. Since our
calculation includes dileptons from $\pi^+ \pi^-$ resonant production as well
as $\pi \rho$ scattering, adjustments should be made to compensate for possible
double counting. Charged pions that annihilate to form a lepton pair cannot also
scatter with a neutral rho to form a lepton pair and a pion. This effect has
been accounted for in our results. It should be
noted that any effects arising from the lifetime of pions and rhos have not 
been accounted for, as it is
possible for the pions and rhos to decay inside the reaction zone before having
a chance to interact. In the final addition made to our simulation, we allow 
the original projectile
nucleons that didn't participate in the primary scattering stage to 
scatter with pions in the secondary (intermediate) stage. These unscattered
projectile nucleons could contribute significantly to the overall multiplicity,
provided that they exist in sufficient numbers. Based on the largest region of
excess in the lepton pair spectra, we focused on the production of the $\eta$
meson by this mechanism. This type of 
rescattering is handled
entirely by PYTHIA.  

The total dilepton yield from our model is the sum of lepton pairs from primary
plus secondary scattering. The invariant mass distributions of 
the dileptons from all contributions will be discussed in the last section.

A reasonable candidate for a successful model description of ultra-relativistic
heavy-ion collisions must at minimum be able to reproduce the rapidity
distributions and transverse spectra of pions produced in the collisions. We
have performed these tests with our model and compared the results to available
experimental data at CERN\cite{me1}. We will not repeat this analysis here, but
only state the results: the total number of produced pions is reproduced to
better than a factor of two; the shape of the rapidity distribution shows the
correct degree of stopping; the slope of the transverse momenta is reproduced. 
It should be noted that our model was also tested against the proton-induced
interactions (p+Be and p+Au) at CERN\cite{me1}. In both cases, our simulation
reproduced the lepton pair data by only considering the primary decays of
pseudoscalar and vector mesons. It is also reassuring that the dilepton spectra
from these decays in our model were consistent with the cocktail predictions
made by the CERES collaboration for all systems: p+Be, p+Au, S+Au, and Pb+Au.

With the inclusion of the secondary scattering described in the 
previous section, the invariant mass distributions of dileptons are 
shown in Fig.\~1 for S+Au and Fig.\~2 for Pb+Au.  Both systems have a 
marked increase in lepton-pair production between an invariant mass of 
200 and 500 MeV as well as a noticeable increase in the higher mass 
region when compared with the spectra without secondary scattering 
included.  The bulk of the increase is attributed to non-resonant 
$\pi$ $\rho$ scattering, not pseudoscalar and vector meson production 
by secondary scattering of projectile nucleons with pions.  The latter 
effect is minimal, as on average only about 20 (5) projectile nucleons 
rescatter with pions in the Pb- (S-) induced collisions.  The 
proton-induced reactions do not show any significant increase in 
dilepton production with secondary scattering included.  It is not 
surprising that secondary scattering becomes important in the 
nucleus-nucleus systems, as a denser nuclear medium is created during 
these collisions as compared to the proton-induced collisions.

When comparing the differences between our calculations and the 
experimental results for the S+Au dilepton spectrum to those found in Pb+Au,
one finds that our calculations are much closer to experiment for the 
heavier projectile.  Since the difference between conventional theory and 
experiment is presumed to be due to in-medium vector meson mass shifts \cite{rhoshift}, 
this tendency is puzzling.

Conventional methods, such as the rescattering studied in this paper, cannot
account for the entire excess of electron-positron pairs found in either the
S+Au or Pb+Au systems. Despite this fact, the estimates made in these studies
indicate that a true understanding of the dilepton mass spectra at CERN SPS
should include the rescattering effects investigated here.

\begin{figure}[hbt]
\includegraphics{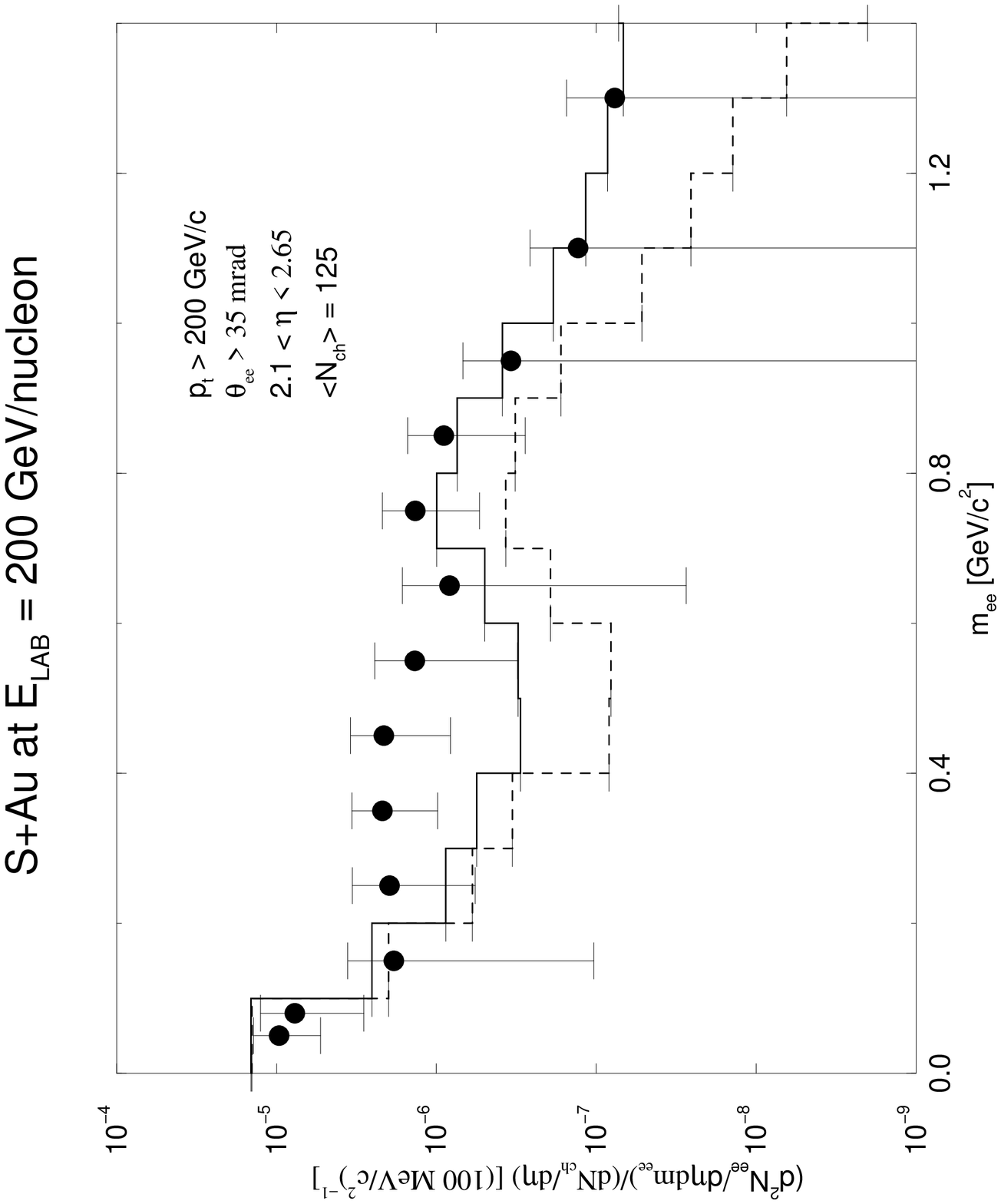}
\vspace*{14cm}
\caption{Total dilepton invariant mass distributions: primary meson decays
alone (dashed line) and primary meson decays with 
secondary scattering (solid line) as compared with CERES data for S+Au
collisions.\label{sau1}}
\end{figure}
\pagebreak

\begin{figure}[hbt]
\includegraphics{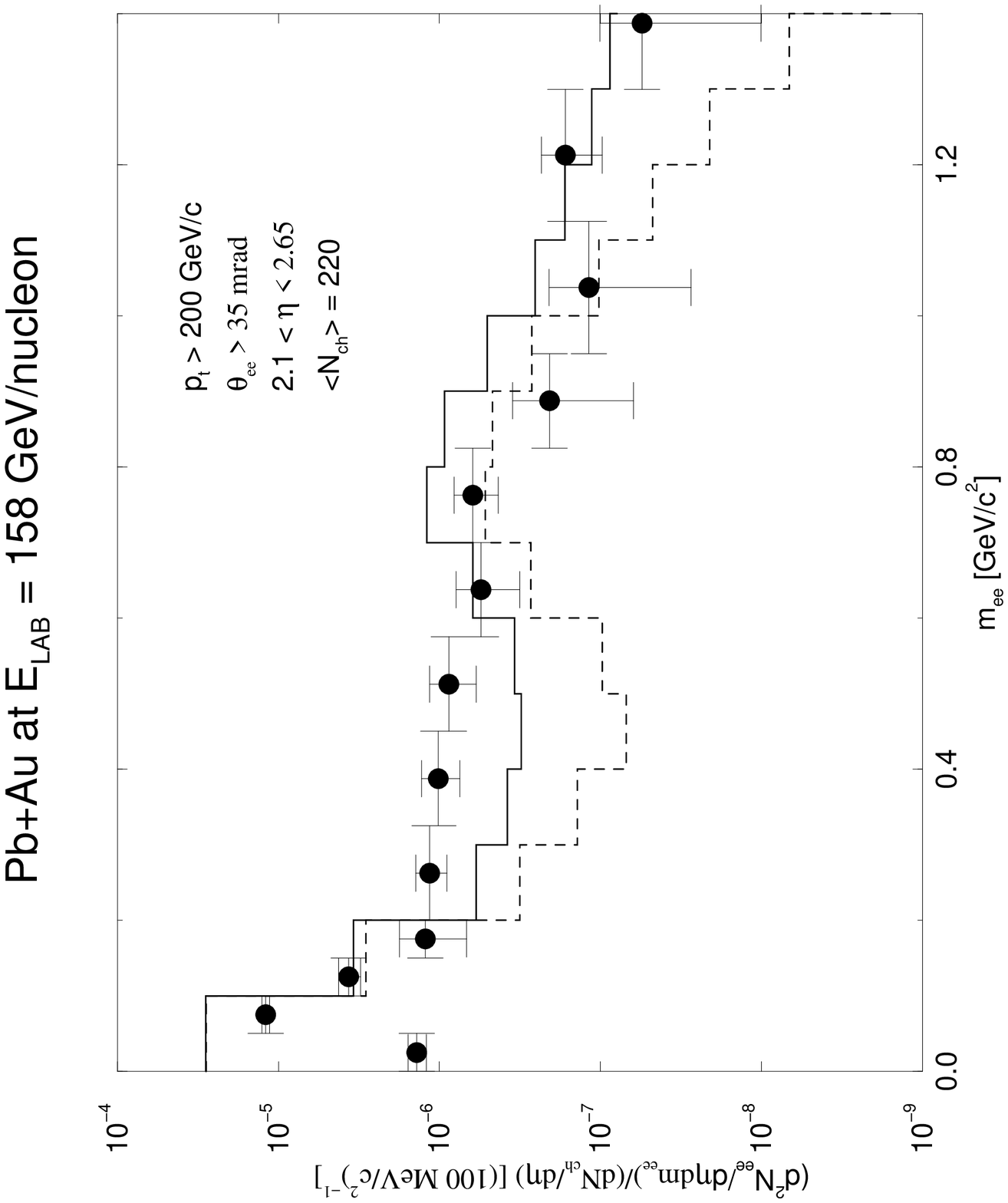}
\vspace*{14cm}
\caption{Total dilepton invariant mass distributions: primary meson decays
alone (dashed line) and primary meson decays with 
secondary scattering (solid line) as compared with CERES data for Pb+Au
collisions.\label{pbau3}}
\end{figure}

\end{document}